\documentclass[%
 reprint,
superscriptaddress,     
 amsmath,amssymb,
 aps,
]{revtex4-2}

\usepackage{graphicx}
\usepackage{dcolumn}
\usepackage{bm}
\usepackage{balance}

\usepackage{float} 

\begin{document}

\preprint{APS/123-QED}

\title{Dependence of the Nonlinear-Optical Response of Materials on their Linear $\epsilon$ and $\mu$}


\author{Diego M. Sol\'{i}s}
\affiliation{Department of Electrical and Systems Engineering, University of Pennsylvania, Philadelphia, Pennsylvania, USA}
\author{Robert W. Boyd}
\affiliation{Department of Physics, University of Ottawa, Ottawa, Ontario, Canada}
\affiliation{Institute of Optics and Department of Physics and Astronomy, University of Rochester, Rochester, New York, USA}
\author{Nader Engheta}
\affiliation{Department of Electrical and Systems Engineering, University of Pennsylvania, Philadelphia, Pennsylvania, USA}

\date{\today}

\begin{abstract}
We investigate, theoretically and numerically, the dependence of a material's nonlinear-optical response on the linear relative electric permittivity $\epsilon$ and magnetic permeability $\mu$. The conversion efficiency of low-order harmonic-generation processes, as well as the increase rate of Kerr-effect nonlinear phase shift and nonlinear losses from two-photon absorption (TPA), are seen to increase with decreasing $\epsilon$ and/or increasing $\mu$. We also discuss the rationale and physical insights behind this nonlinear response, particularly its enhancement in $\epsilon$-near-zero (ENZ) media. This behavior is consistent with the experimental observation of intriguingly high effective nonlinear refractive index in degenerate semiconductors such as indium tin oxide [\textit{Alam et al., Science 352 (795), 2016}] (where the nonlinearity is attributed to a modification of the energy distribution of conduction-band electrons due to laser-induced electron heating) and aluminum zinc oxide [\textit{Caspani et al., Phys. Rev. Lett. 116 (233901), 2016}] at frequencies with vanishing real part of the linear permittivity. Such strong nonlinear response can pave the way for a new paradigm in nonlinear optics with much higher conversion efficiencies and therefore better miniaturization capabilities and power requirements for next-generation integrated nanophotonics.
\end{abstract}

\maketitle


\section{\label{sec:level1}Introduction}

Many research endeavors have focused on the quest for materials with strong and fast nonlinear light-matter interactions. Large ultrafast nonlinear optical responses are paramount for a plethora of applications relying on active photonic integrated circuits, ranging from all-optical signal processing \cite{Cotter1523,Vo:10} to quantum computers \cite{PhysRevLett.75.4337,Leach662}. But the integration density of these devices, if based on nonresonant nonlinear processes (hinging upon virtual transitions and ergo very fast), is burdened by the intrinsic perturbative nature of such nonlinear phenomena, which typically require high optical intensities and/or long interaction lengths. In order to circumvent this weak response, diverse alternatives have been proposed, aimed at extrinsically boosting nonlinearities with tailored electromagnetic resonances by means of structuring materials, like micro-cavities \cite{Bravo-Abad:07}, slow-light photonic-crystal waveguides \cite{PhysRevLett.87.253902,SoljaCiC2004}, metallo-dielectric composites \cite{PhysRevLett.93.123902}, or plasmonic nanostructures \cite{Wurtz2011,Cai1720,Kauranen2012}.

Moreover, materials with near-vanishing permittivity, known as $\epsilon$-near-zero (ENZ) materials, were initially predicted \cite{PhysRevA.81.043839,PhysRevB.85.045129} (by virtue of either electric field enhancement or better phase-matching) and later observed \cite{Suchowski1223,doi:10.1063/1.4917457,Kinsey:15,doi:10.1021/acsphotonics.5b00355,Alam795,PhysRevLett.116.233901,Alam2018} to enhance nonlinear processes. More recently, transparent conductive oxides such as indium tin oxide (ITO) and Al-doped ZnO (AZO) have drawn much attention as promising candidates to increase the strength of nonlinear interactions. These degenerately doped semiconductors (i) are complementary metal-oxide-semiconductor (CMOS)-compatible, (ii) have an ENZ wavelength in the near-IR (tuned by varying post-deposition annealing time and temperature) for which the nonlinear refractive index has been experimentally measured to be unprecedentedly large \cite{Alam795,PhysRevLett.116.233901}---up to several orders of magnitude larger than the previously reported largest value (As$_2$Se$_3$ glass) \cite{Eggleton2011}---and with a sub-picosecond response time, and (iii) provide less loss than noble metals in this spectral region. In fact, the nonlinear response of these materials is so large that one might question whether the usual expansion of the material polarization as a power series in electric field \cite{Boyd:2008:NOT:1817101} is still valid. As pointed out in \cite{Reshef:17}, there may still be a convergent power series for the polarization in terms of the electric field amplitude in this regime, although the widely used expression for the intensity-dependent refractive index \cite{Boyd:2008:NOT:1817101} $n$=$n_0\!+\!n_2I$ ($n_0$ being the linear refractive index, $n_2$=$\frac{3\chi^{(3)}}{4n_0\text{Re}\left\{n_0\right\}\epsilon_0c}$ the nonlinear coefficient, $\chi^{(3)}$ the third-order nonlinear susceptibility, and $I$ the optical intensity), stems from a Taylor expansion that under ENZ conditions is divergent and should therefore be reassessed. Thus the dependence of $n$ on $I$ is non-perturbative, even though the dependence of the polarization on field strength remains perturbative.

In this manuscript, the theoretical analysis of wave propagation in a nonlinear medium with second- or third-harmonic, instantaneous (nondispersive) susceptibilities is revisited, and the dependence of the nonlinear response on the linear part of the relative dielectric permittivity $\epsilon$, which is allowed to be dispersive, is studied in detail. For the sake of completeness, the variation of linear relative magnetic permeability $\mu$ is also taken into account. Furthermore, a finite-difference time-domain (FDTD) \cite{taflove2005computational} full-wave electromagnetic solver has been implemented (generalized for dispersive media and for arbitrary nonlinear phenomena) to validate the theoretical predictions. It will be shown that phase-matched nonlinear propagation has a conversion efficiency that tends to increase with decreasing $\epsilon$ and/or increasing $\mu$, because the inverse of the conversion length tends to increase with an increasing relative impedance $\eta$=$\sqrt{\mu/\epsilon}$. Additionally, the intensity of the reflected second/third harmonics tends to increase with decreasing $\epsilon$ and/or $\mu$ for normal incidence from vacuum to a semi-infinite region of such nonlinear media. When phase-mismatch is brought into play, it is well-known that destructive interference inhibits the harmonic conversion process and a characteristic space-periodic pattern shows up; it will be shown that the maxima of these periodic oscillations either increase with $\mu$ and/or $1/\epsilon$, or remain constant but with a spatial frequency that is roughly proportional to the same factor $\sqrt{\mu/\epsilon}$, so the effective conversion length is reduced as $\epsilon$ ($\mu$) decreases (increases). We will also connect this $\eta$-dependence observed in harmonic-generation processes with the fact that the second-order index of refraction and the two-photon absorption (TPA) coefficient increase with increasing $\eta$ as well.

\section{Theory and Numerical Results}
\newcommand{\comment}[1]{}
For simplicity and without loss of generality, let us focus our description (we herein extend the analytical framework in \cite{PhysRev.127.1918,Boyd:2008:NOT:1817101} to include the effect of linear magnetic permeability) on second-harmonic generation within a medium that is lossless at the fundamental and second-harmonic frequencies, $\omega_1$ and $\omega_2$=$2\omega_1$, respectively. We consider plane-wave propagation in the $+z$ direction and express $\tilde{E}_j$, the electric field at frequency $\omega_j$ ($j$=1,2), as
\begin{equation}
\tilde{E}_j(z,t)=2\text{Re}\left[E_j(z)e^{-i\omega_jt}\right]=2\text{Re}\left[A_j(z)e^{i(k_jz-\omega_jt)}\right],\label{eq:1}
\end{equation}
where a slowly varying complex amplitude $A_j(z)$ is used, and $k_j$=$n_j\frac{\omega_j}{c}$ is the wavenumber, with $n_j$=$\sqrt{\mu_j\epsilon_j}$ the refractive index, $\mu_j$ and $\epsilon_j$  being the linear relative permeability and permittivity at frequency $\omega_j$, respectively. The presence of nonlinear polarization $\tilde{P}_{NL,j}$ leads to the following well-known inhomogeneous wave equation for $\tilde{E}_j$ \cite{Boyd:2008:NOT:1817101}:
\begin{equation}
\frac{\partial ^2\tilde{E}_j(z,t)}{\partial z^2}-\frac{\mu_j\epsilon_j}{c^2}\frac{\partial ^2\tilde{E}_j(z,t)}{\partial t^2}=\frac{\mu_j}{c^2 \epsilon _0}\frac{\partial ^2\tilde{P}_{NL,j}(z,t)}{\partial t^2},\label{eq:2}
\end{equation}
with:
\begin{subequations}
\begin{gather}
\tilde{P}_{NL,1}(z,t)=2\text{Re}\left[2\epsilon_0A_2(z)A_1^*(z)\chi^{(2)}e^{i((k_2-k_1)z-\omega_1t)}\right],\label{eq:3a}\\
\tilde{P}_{NL,2}(z,t)=2\text{Re}\left[\epsilon_0A_1^2(z)\chi^{(2)}e^{i(2k_1z-\omega_2t)}\right],\label{eq:3b}
\end{gather}\label{eq:3}%
\end{subequations}
where the second-order nonlinear optical susceptibility is denoted by $\chi^{(2)}$. By placing Eqs.~(\ref{eq:1}) and (\ref{eq:3}) into Eq.~(\ref{eq:2}), and making the slowly varying amplitude approximation (SVAA), $\vert\frac{d^2A_j}{dz^2}\vert \ll \vert k_j\frac{dA_j}{dz}\vert$, it is straightforward to arrive at the pair of coupled-amplitude equations:
\begin{subequations}
\begin{gather}
\frac{dA_1}{dz}=i\frac{\eta_1\omega_1\chi^{(2)}}{c}A_2(z)A_1^*(z)e^{-i\Delta k z},\label{eq:4a}\\
\frac{dA_2}{dz}=i\frac{\eta_2\omega_2\chi^{(2)}}{2c}A_1^2(z)e^{i\Delta k z},\label{eq:4b}
\end{gather}\label{eq:4}%
\end{subequations}
where $\Delta k$=$2k_1\!-\!k_2$. It is convenient to introduce normalized field amplitudes $u_j(z)$=$\sqrt{I_j(z)/I}$, where $I_j(z)$=$\frac{2\vert A_j(z)\vert^2}{\eta_0\eta_j}$ is the intensity of the $j$-th harmonic, $\eta_0$=$\sqrt{\mu_0/\epsilon_0}$ is the intrinsic impedance of vacuum and $\eta_j=\sqrt{\mu_j/\epsilon_j}$ is the relative impedance. Following the Manley-Rowe relations, the total intensity $I$ is constant, so $\Sigma_j u_j^2=1$. If we define a characteristic interaction length
\begin{equation}
l=\frac{c}{\omega_1\chi^{(2)}}\sqrt{\frac{2}{\eta_1^2\eta_2\eta_0I}},\label{eq:5}
\end{equation}
a measure of the normalized phase velocity mismatch will be $\Delta s$=$\Delta kl$.

\subsection{Perfect Phase-Matching}
\begin{figure}[h]
\includegraphics[width=3.4in]{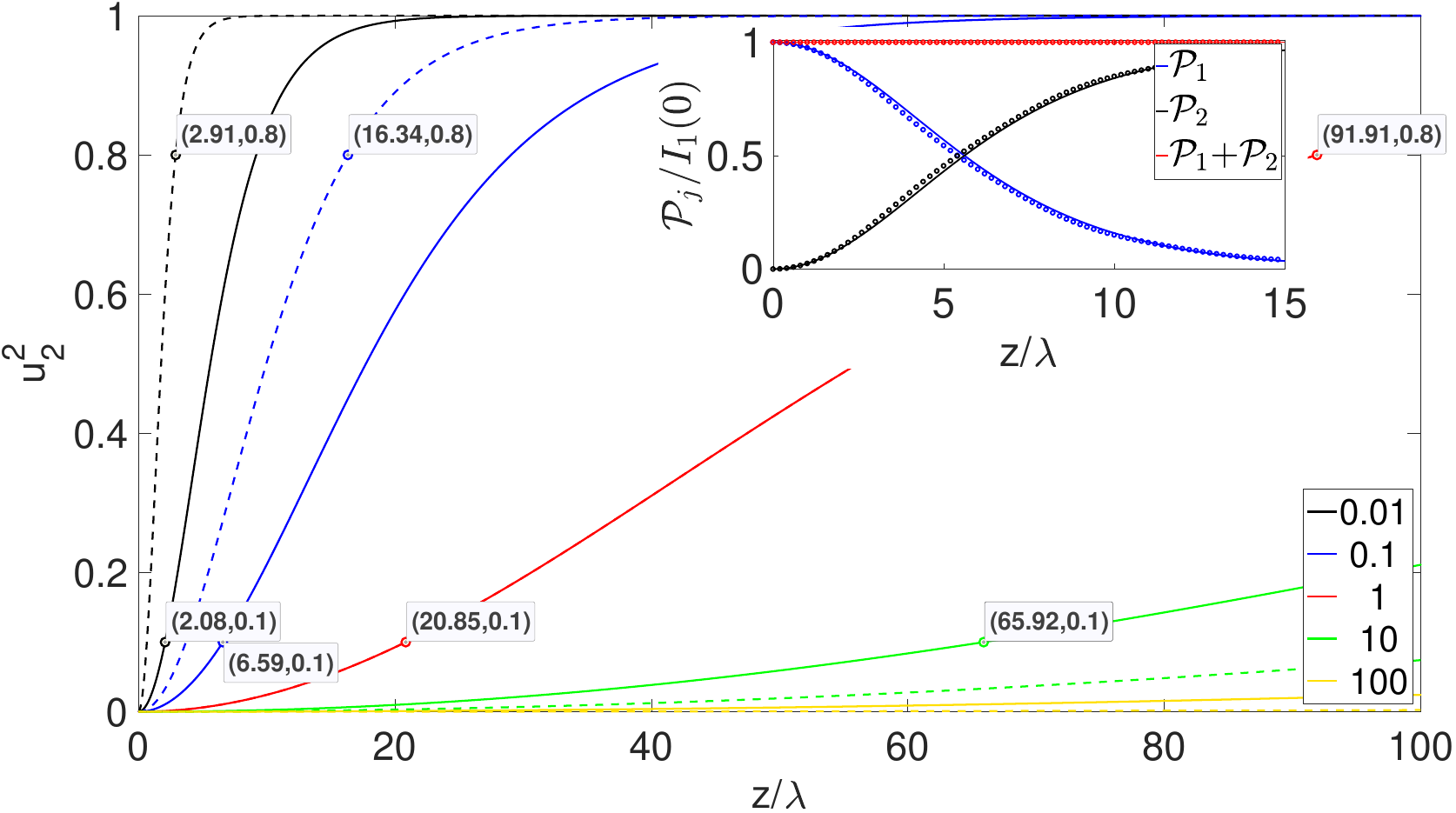}
\caption{Normalized intensity of the second-harmonic wave as predicted by the analytic solution of Eq.~(\ref{eq:8}) vs. $z/\lambda$ for various values of $\epsilon$ and for $\chi^{(2)}$=$5\!\times\!10^{-12}$ [m/V]. Solid lines: fixed $A_1(0)$=$5\!\times\!10^{8}$ [V/m] (the data cursors give the values $(z/\lambda,u_2^2)$ at the indicated point and mark the distances at which $u_2^2(z)$=0.1, illustrating the previously mentioned $\eta$ $z$-scaling with $\eta$). Dashed lines: fixed $I$=$2(5\times10^{8})^2/\eta_0$, i.e., the intensity corresponding to the previous fixed value of $A_1(0)$ when $\epsilon_1$=1 (the data cursors mark the distances at which $u_2^2(z)$=0.8, illustrating now the $\sqrt{\eta^3}$ $z$-scaling with $\eta$). The inset displays, for $\epsilon$=0.01, the time-averaged Poynting vector $\mathcal{P}_j $ with (solid lines) and without (markers) the SVAA approximation, showing that the SVAA is almost perfectly valid. In this latter case, the initial condition $d^2A_2(0)/dz^2$=0 is applied. Note that $\mathcal{P}_j/I_1(0)$ is equal to $u_j^2$ only when the SVAA approximation is considered, so only $\mathcal{P}_j$, and not $u_j^2$ as defined so far, can be compared.}
\label{fig:1}
\end{figure}

If the phase velocity of both harmonics is the same, we have $\Delta s$=0. In this scenario, 
one can make use of the fact that $u_1(z)^2u_2(z)\text{cos}\big(\theta(z)\big)$ is a conserved quantity \cite{PhysRev.127.1918,Boyd:2008:NOT:1817101}, with $\theta$=$2\phi_1(z)-\phi_2(z)+\Delta kz$ ($\phi_j$ being the phase of the complex amplitude $A_j$), and use $\Gamma$=$u_1(0)^2u_2(0)\text{cos}\big(\theta(0)\big)$ to decouple Eqs.~(\ref{eq:4a}),(\ref{eq:4b}). After some lengthy mathematical manipulations, and using $\zeta$=$z/l$ \cite{PhysRev.127.1918,Boyd:2008:NOT:1817101}, one can arrive at an equation expressed only in terms of $u_2(\zeta)$
\begin{equation}
\frac{d^2u_2^2(\zeta)}{d\zeta^2}=2\pm\sqrt{u_2^2(\zeta)\big(1-u_2^2(\zeta)\big)^2-\Gamma^2},\label{eq:6}
\end{equation}
whose general solution has the form of the elliptic integral
\begin{equation}
\zeta=\pm\frac{1}{2} \int_{u_2(0)}^{u_2}\frac{d(u_2^2)}{\sqrt{u_2^2(1-u_2^2)^2-\Gamma^2}}.\label{eq:7}
\end{equation}
$u_2^2$, which oscillates between the two lowest positive roots of the integrand's denominator, can thus be expressed in closed form with the help of the Jacobi elliptic function $sn()$ \cite{whittaker1996course}. Nonetheless, assuming $u_2(0)$=0, i.e., only the fundamental frequency impinges on the semi-infinitely extended nonlinear medium, the solution is reduced to the simpler form:
\begin{equation}
u_1(\zeta)=\text{sech}(\zeta),\;\;\;\;\;\;u_2(\zeta)=\text{tanh}(\zeta).\label{eq:8}
\end{equation}
In terms of these results we can immediately find the intensity and amplitude conversion efficiencies from the $\omega_1$ wave to the $\omega_2$ wave, defined as $\frac{I_2(z)}{I_1(0)}$ and $\frac{\vert A_2(z)\vert}{\vert A_1(0)\vert}$ respectively, as $u_2^2(z)$ and $u_2(z)$. From inspection of Eq.~(\ref{eq:8}), it is thus clear that conversion efficiency increases with increasing $l^{-1}$, which will vary with $\sqrt{\eta_1^2\eta_2I_1(0)}$ or, equivalently, with $\sqrt{\eta_1\eta_2}\vert A_1(0)\vert$. For the perfect phase-matching condition in an isotropic medium, we need to have $2\omega_1\sqrt{\mu_1\epsilon_1}\!=\!\omega_2\sqrt{\mu_2\epsilon_2}$. This can be achieved in several different ways: (1) For non-magnetic isotropic materials where $\mu_1\!=\!\mu_2\!=\!1$, phase matching occurs when $\epsilon_1\!=\!\epsilon_2$, which is possible when we are far away from any resonance of the material and $\epsilon_1\!=\!\epsilon_2\!>\!1$. However, near the ENZ frequencies, the permittivity function is dispersive and thus it should be a function of frequency. Therefore, the condition $\epsilon_1\!=\!\epsilon_2$ can be achieved near zero crossing of the dispersion curves at $\omega_1$ and $\omega_2$ with properly engineered materials with two or more Lorentzian dispersions (or one Drude and one or more Lorentzian dispersions); (2) for the case of magnetic isotropic materials, we can have $\epsilon_1\!\neq\!\epsilon_2$ when $\mu_1\!\neq\!\mu_2$ such that $2\omega_1\sqrt{\mu_1\epsilon_1}\!=\!\omega_2\sqrt{\mu_2\epsilon_2}$. There are other cases such as anisotropic materials in which the phase-matching condition may occur for a given direction of propagation. Here, for the sake of simplicity, we assume the first case. When $\epsilon$=$\epsilon_1$=$\epsilon_2$ and $\mu$=$\mu_1$=$\mu_2$, we have a stretching/compression of the $z$-axis by a factor $h$ such that $u_{2,\eta=h}(z)$=$u_{2,\eta=1}(hz)$ when $A_1(0)$ is fixed, or $u_{2,\eta=h}(z)$=$u_{2,\eta=1}(\sqrt{h^3} z)$ when $I$=$I_1(0)$ is fixed (moreover, for small $\zeta$, given that $\text{tanh}(\zeta)\!\approx\!\zeta$, the intensity and amplitude conversion efficiencies scale with $\eta^3$ and $\eta$, respectively). This behavior can be seen in Fig.~\ref{fig:1}, which shows the evolution of $u_2^2$ vs. distance (normalized with respect to the wavelength of the fundamental frequency in vacuum $\lambda$) for $\chi^{(2)}$=$5\!\times\!10^{-12}$ [m/V] and different values of $\epsilon$=$\epsilon_1$=$\epsilon_2$ ranging from 0.01 to 100 ($\mu$=$\mu_1$=$\mu_2$ is set to 1), while keeping the electric field amplitude constant (solid lines) or the intensity constant (dashed lines). One can see that identical curves would be obtained by setting $\epsilon$=1 and varying $\mu$ from 100 to 0.01. Moreover, the magnetic permeability can be used as an extra degree of freedom to achieve phase-matching, by choosing the permittivities and permeabilities such that $\mu_1\epsilon_1$=$\mu_2\epsilon_2$. Crucially, we note that the SVAA approximation loses its validity as $\epsilon\mu$ is reduced (the wavelength increases and thus the term $\vert k_j\frac{dA_j}{dz}\vert$ decreases), which in Fig.~\ref{fig:1} especially concerns the case for which $\epsilon$=0.01. A numerical resolution of the two coupled equations resulting from adding the terms $\frac{d^2A_j}{dz^2}$ to Eqs.~(\ref{eq:4a}),(\ref{eq:4b}) yields, however, time-averaged Poynting vector curves that are very close to the ones obtained with the second of Eqs.~(\ref{eq:8}) (see inset in Fig.~\ref{fig:1}).

Analagous derivations for third-harmonic generation (considering nonlinear processes characterized by $\chi^{(3)}(3\omega;\omega,\omega,\omega)$ and $\chi^{(3)}(\omega;3\omega,-\omega,-\omega)$) will yield, in the SVAA approximation and considering $\Delta k$=$0$ and $u_3(0)$=$0$, a closed-form solution of the form
\begin{equation}
u_1(\zeta)=\frac{1}{\sqrt{1+\zeta^2}},\;\;\;\;\;\;u_3(\zeta)=\frac{\zeta}{\sqrt{1+\zeta^2}},\label{eq:9}
\end{equation}
where the characteristic interaction length $l$=$z/\zeta$ is now defined as:
\begin{equation}
l=\frac{c}{3\omega_1\chi^{(3)}}\frac{4}{\sqrt{\eta_1^3\eta_3}\eta_0I}.\label{eq:10}
\end{equation}
The distance scale for conversion of the fundamental to the third harmonic will thus decrease as $\sqrt{\eta_1^3\eta_3}I$ or, equivalently, as $\sqrt{\eta_1\eta_3}\vert A_1(0)\vert^2$. Assuming $\epsilon$=$\epsilon_1$=$\epsilon_3$ and $\mu$=$\mu_1$=$\mu_3$, $u_{3,\eta=h}(z)$=$u_{3,\eta=1}(hz)$ or $u_{3,\eta=h}(z)$=$u_{3,\eta=1}(h^2 z)$ will hold when either $A_1(0)$ or $I$ are fixed, respectively. For small $\zeta$, the intensity and amplitude conversion efficiencies thus scale with $\eta^4$ and $\eta$, respectively.
\begin{figure}[h!]
\centering
\includegraphics[width=3.4in]{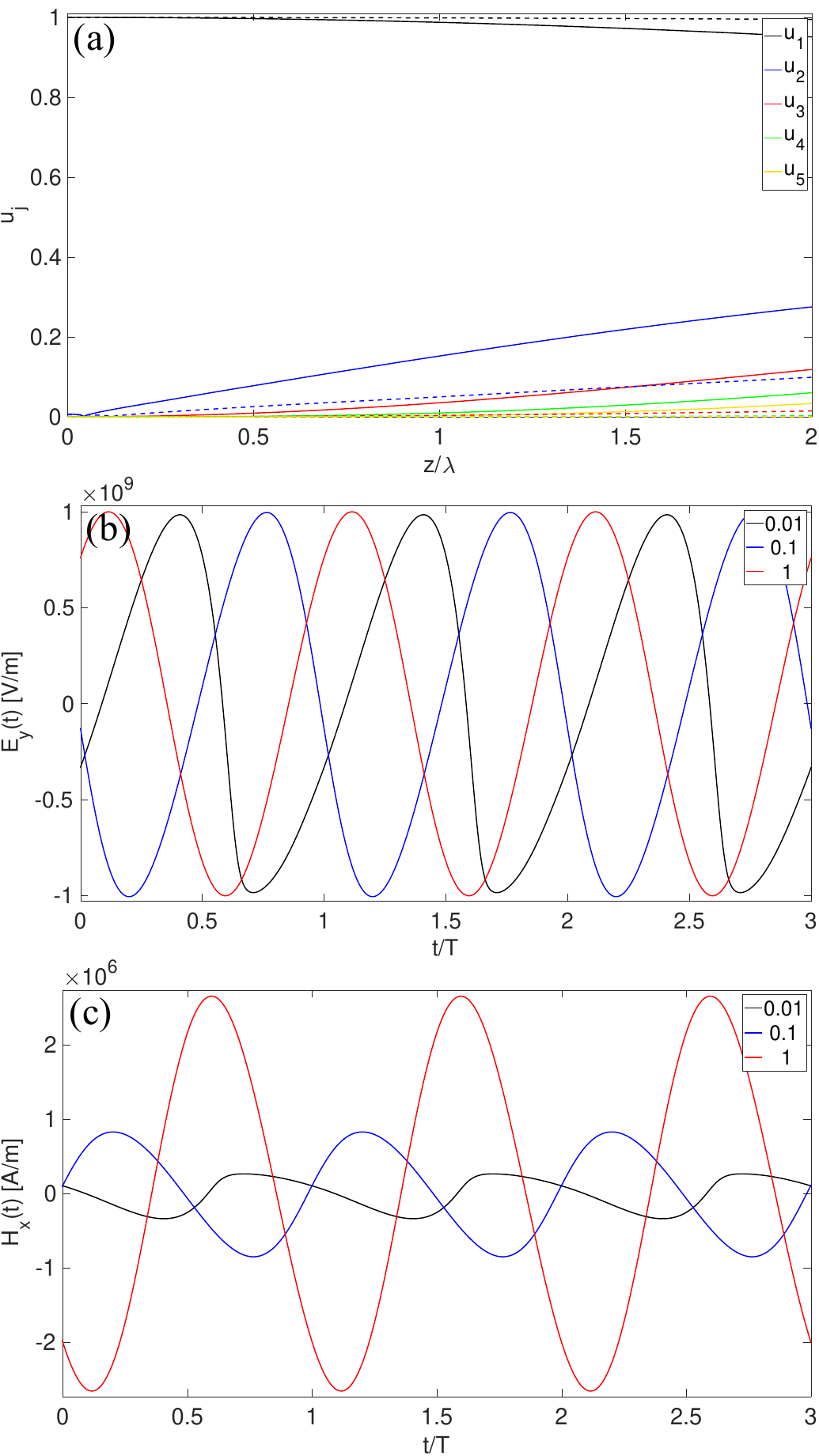}
\caption{Second-order nonlinear response obtained through FDTD simulations. (a) Normalized field amplitudes of the first five harmonics for $\epsilon$=0.01 and 0.1 (solid and dashed lines, respectively) vs. depth normalized to the vacuum wavelength into the nonlinear medium. (b),(c) Temporal variation of the electric and magnetic fields at a distance of two vacuum wavelengths of the fundamental wave from the interface, for different values of $\epsilon$. Comparison of panels (b) and (c) seems to suggest that $H_x$ displays more distortion than $E_y$, which is consistent with the additional distortion coming from the term $\frac{dA_{y,j}(z)}{dz}$ in $H_x(z,\omega_j)$=$-\frac{1}{i\omega_j\mu_0\mu}\left(ik_jA_{y,j}(z)+\frac{dA_{y,j}(z)}{dz}\right)e^{ik_jz}$. However, this augmented distortion in $H_x$ is much less pronounced than it seems from visual inspection, and is mostly owed to the ratio of the horizontal and vertical  axes (i.e., if both $E_y$ and $H_x$ are normalized to 1, their distortions look very similar).}
\label{fig:2}
\end{figure}

Let us now study the totality of nonlinear processes arising from such instantaneous (nondispersive) second-order $\big(P^{(2)}(t)$=$\epsilon_0\chi^{(2)}E^2(t)\big)$ and third-order---or Kerr effect--- $\big(P^{(3)}(t)$=$\epsilon_0\chi^{(3)}E^3(t)\big)$ nonlinear polarizations. In order to do so, we developed an FDTD algorithm \cite{558652,Dissanayake:10} incorporating these nonlinear interactions, which naturally cover the entire optical spectrum when described in the time domain (that is, all higher harmonics and their nonlinear interactions are implicitly taken into account). We consider a ``half-space'' problem (effectively achieved with perfectly matched layers) where an incident plane-wave in vacuum meets the interface with the nonlinear medium with an electric field amplitude normalized such that, regardless of the different values of $\epsilon$ considered (from 0.01 to 1), the transmitted electric field immediately on the other side of the interface is kept constant and equal to $E_0$=$2A_1(0)$=$10^{9}$ [V/m]. Of course, strictly speaking, higher harmonics may already be created in reflection at the interface, so the initial condition $u_{2,3}(0)$=0 assumed in our previous analysis is not, in general, strictly satisfied here anymore \cite{PhysRev.128.606,Boyd:2008:NOT:1817101} (see also Eqs.~(\ref{eq:16}),(\ref{eq:17})); depending on the initial phase difference among these harmonics, they may actually first decrease to zero before steadily increasing.

\begin{figure}[h]
\centering
\includegraphics[width=3.4in]{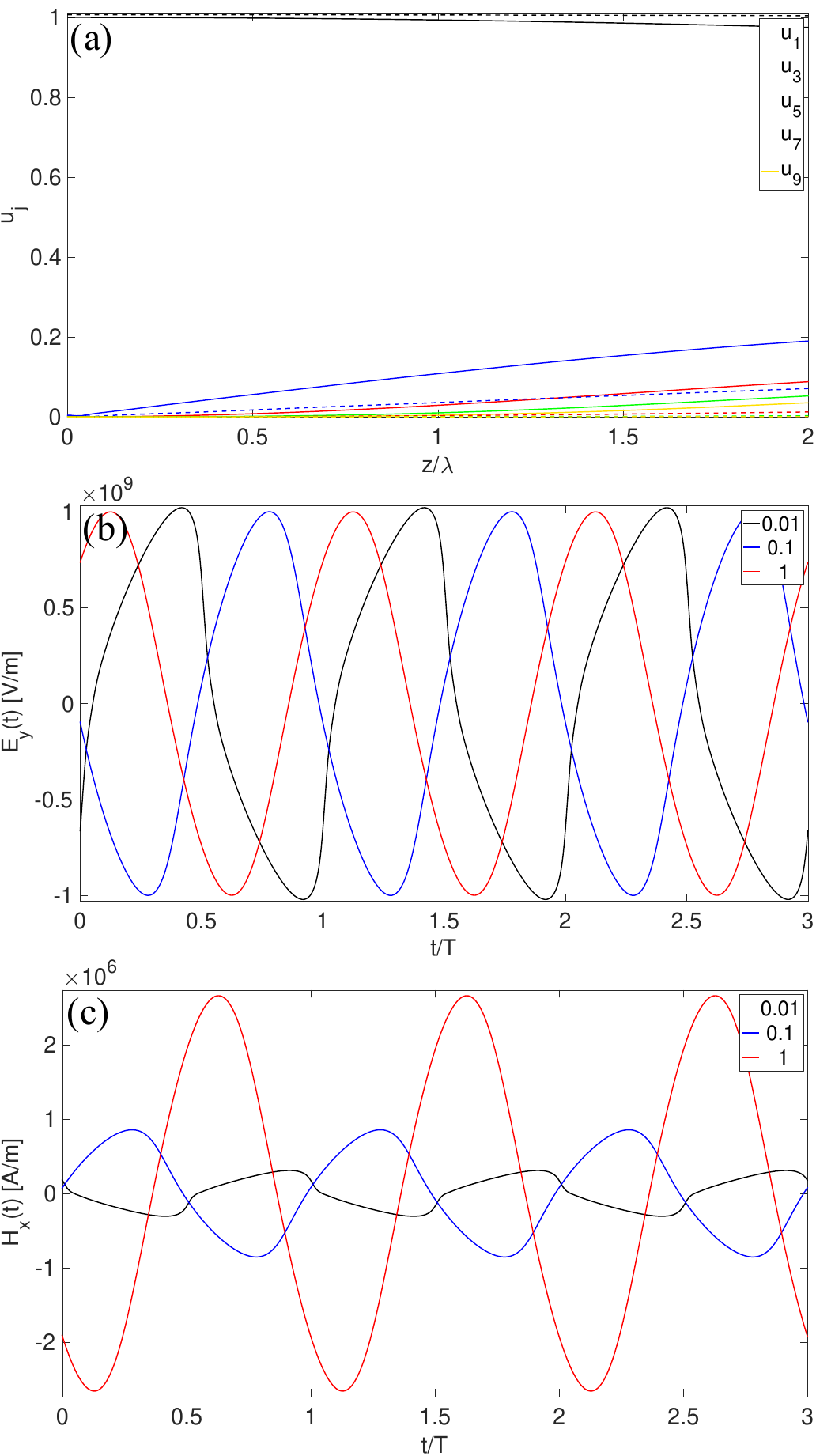}
\caption{Third-order nonlinear response calculated numerically with FDTD. (a),(b),(c) Same as in Fig.~\ref{fig:2}, but with $\chi^{(3)}\!=\!\chi^{(2)}/E_0$=$5\!\times\!10^{-21}$ [(m/V)$^2$].}
\label{fig:3}
\end{figure}

The numerical results in Figs.~\ref{fig:2},\ref{fig:3}, obtained from FDTD simulations, very clearly depict the increase of nonlinear response as $\epsilon$ is reduced, in agreement with the analytical analysis shown above. It is also interesting to point out that, in normalizing the electric field transmitted through the interface, the transmitted intensity actually decreases with $\sqrt{\epsilon}$. Therefore, if one normalizes transmitted intensity rather than amplitude, the distances in Fig.~\ref{fig:2}a (Fig.~\ref{fig:3}a) will be $1/\sqrt[4]{\epsilon}$ ($1/\sqrt{\epsilon}$) times shorter. This is consistent with the conversion efficiencies previously predicted by our analytic model. Figs.~\ref{fig:2}c,\ref{fig:3}c depicting the magnetic field are the most revealing of the underlying physical mechanism explaining this enhancement: for fixed $\vert A_1(0)\vert$, assuming not only $\Delta k$=$0$ but also a nondispersive medium ($\eta_j$=$\eta$, $\forall j$) for simplicity, both second- and third-order processes present a conversion efficiency that increases with increasing $\eta$, that is, a weaker (in relative terms) magnetic field enhances the nonlinear response. Let us gain some more intuitive insight as to why this is the case. Given that nonlocal effects are not under consideration, it is clear that the (local) nonlinear polarization sees its effect ``translated'' from time to space through $\nabla\times\mathbf{H}$, according to Maxwell-Ampère's law. For a $\hat{y}$-polarized plane-wave propagating in $+z$ and keeping the adopted $e^{-i\omega t}$ convention, Maxwell's curl equations can be written as
\begin{subequations}
\begin{gather}
\frac{dH_x(z,\omega_j)}{dz}=-i\omega_j\big(\epsilon_0\epsilon E_y(z,\omega_j)+P_{y,NL}(z,\omega_j)\big),\label{eq:11a}\\
\begin{split}
H_x(z,\omega_j)&=-\frac{1}{i\omega_j\mu_0\mu}\frac{d}{dz}\big(A_{y,j}(z)e^{ik_jz}\big) \\&\approx-\frac{k_j}{\omega_j\mu_0\mu}A_{y,j}(z)e^{ik_jz},\label{eq:11b}
\end{split}
\end{gather}\label{eq:11}%
\end{subequations}
where $\frac{dA_{y,j}(z)}{dz}$ has been neglected in the second form of Eq.~(\ref{eq:11b}). A measure of the effective increase in nonlinear distortion with respect to distance felt by the $j$-th harmonic could be written as
\begin{equation}
\frac{\Bigr\vert\left[\frac{dH_j(z)}{dz}\right]_{NL}\Bigr\vert}{\vert H_j(z)\vert}\approx\frac{\vert \omega_j P_{NL,j}(z)\vert}{\vert \frac{k_j}{\omega_j\mu_0\mu}A_j(z)\vert}=\omega_j\eta_0\eta\frac{\vert P_{NL,j}(z)\vert}{\vert A_j(z)\vert}.\label{eq:12}
\end{equation}
As expected, the factor $\eta$ shows up again. Otherwise, this dependence on $\epsilon$ is consistent with physical intuition in that nonlinear polarization represents a larger fraction of total polarization as $\epsilon$ is reduced. Indeed, Eq.~(\ref{eq:11a}) reveals the contributions of the linear and nonlinear portions of the displacement current, demonstrating that in ENZ media the nonlinear part plays a more dominant role even though the coefficients $\chi^{(2)}$ or $\chi^{(3)}$ are kept unchanged. 
\comment{
Importantly, it is interesting to see how we have arrived to this $\eta$ dependence in the increase rate of nonlinear distortion with a purely classical approach (coupled-mode theory) and, yet, this result is perfectly consistent with a quantum-mechanical interpretation of nonlinear optics. Let us not forget that, ultimately, the origin of harmonic-generation processes is rather quantum mechanical: from this perspective, the nonlinear response will be directly proportional to the number of photons and therefore to intensity as $\frac{I}{\hbar\omega}$=$\frac{2\vert E\vert^2}{\eta_0\eta}$, whereas our classical  In defining the nonlinear polarization as $P_{NL}$ in turn, to intensity photon's energy the nonlinear response has a quantum 
}

\begin{figure}[h]
\centering
\includegraphics[width=3.4in]{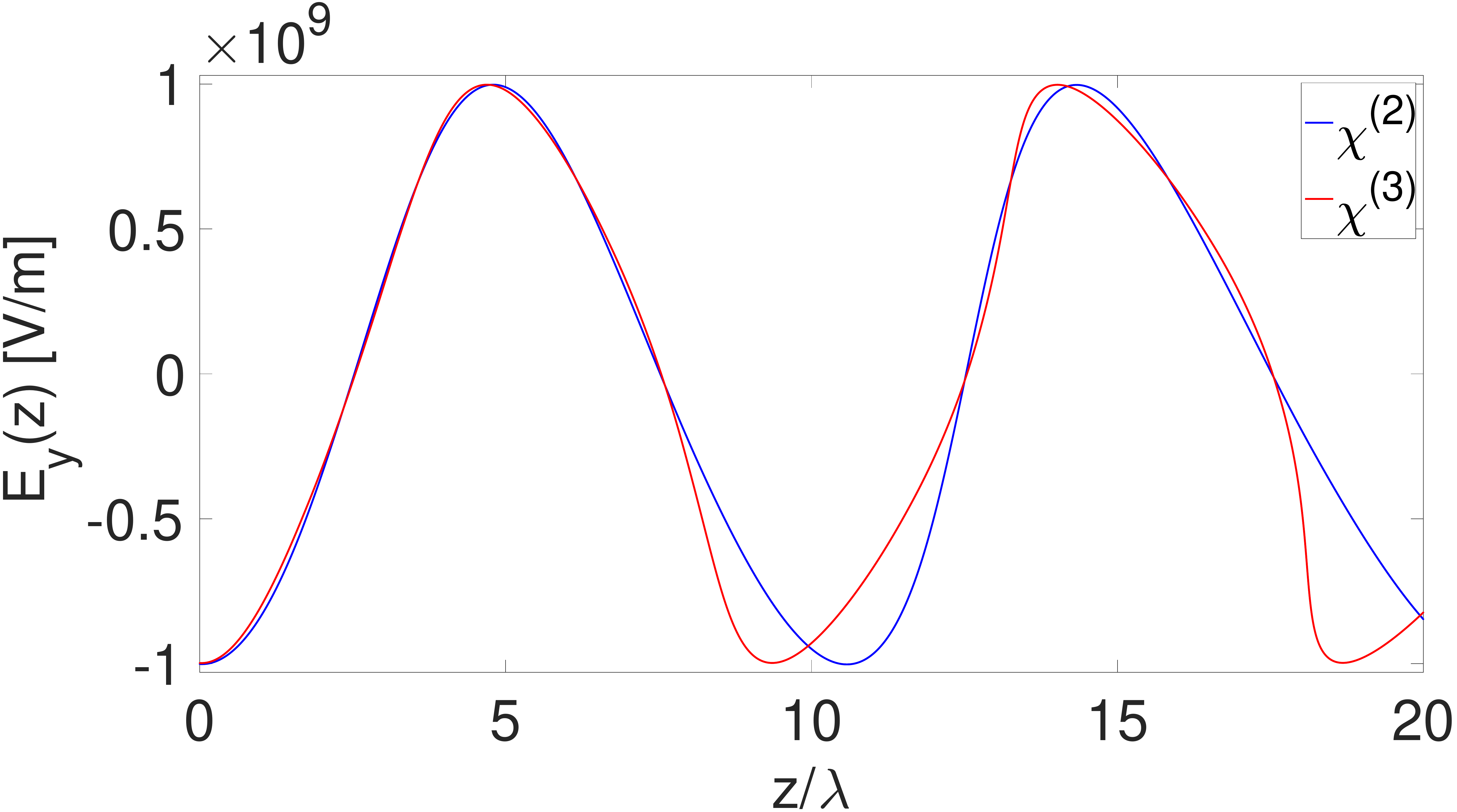}
\caption{FDTD-simulated electric field vs. normalized distance at a given instant in time with $\epsilon$=0.01, for both second- and third-order polarizations.}
\label{fig:4}
\end{figure}

Furthermore, the temporal evolution of the electric field in the case of instantaneous third-order nonlinear polarization, Fig.~\ref{fig:3}b, has intriguing resemblance with the time-inverted version of the so-called relaxation-oscillations \cite{doi:10.1080/14786442608564127} of the Van der Pol nonlinear damped oscillator (well-known in the analysis of circuits containing vacuum tubes), whose oscillation amplitude $x(t)$ obeys the second-order differential equation
\begin{equation}
\frac{d^2x(t)}{dt^2}-\mu\left(1-x^2(t)\right)\frac{dx(t)}{dt}+x(t)=0.\label{eq:13}
\end{equation}
For completeness, a time snapshot of the electric field vs. $z/\lambda$ in Fig.~\ref{fig:4} shows how the waveforms associated with $\chi^{(2)}$ and $\chi^{(3)}$ processes are increasingly distorted with distance when $\epsilon$=0.01. For visualization purposes, given that the wavelength of the fundamental frequency in the nonlinear medium is in this case ten times the vacuum wavelength, we reduce the rate at which distortion increases with $z$ by decreasing the nonlinear susceptibilities by one order of magnitude with respect to Figs.~\ref{fig:2},\ref{fig:3}, and we increase the simulation domain accordingly. It is thought-provoking to see that the wavefront originating from the Kerr effect somewhat reminds us of a shockwave. Actually, one might think of taking advantage of this high spatial-frequency content in highly-resolved microscopy applications.

\subsubsection{Intensity-dependent refractive index}

\begin{figure}[H]
\centering
\includegraphics[width=3.4in]{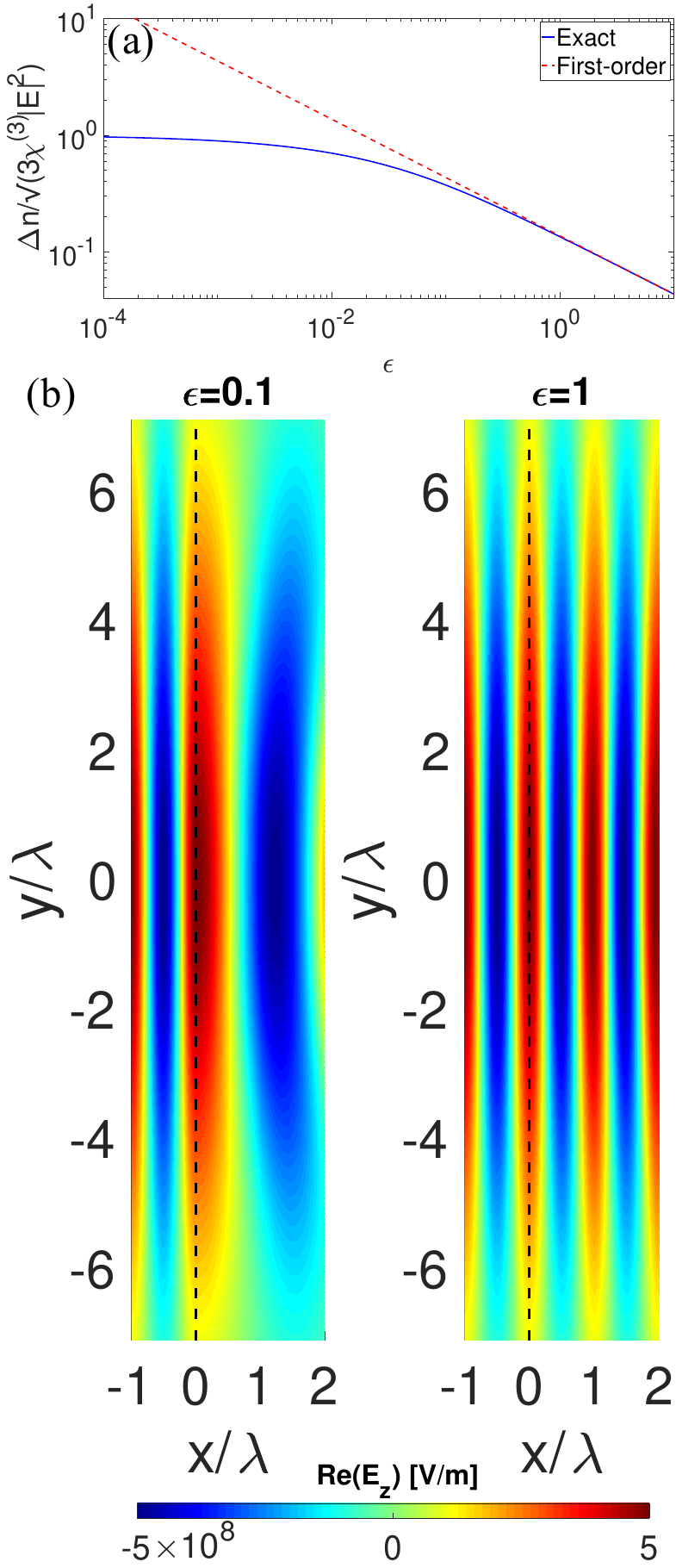}
\caption{(a) $\Delta n$ (normalized with respect to its asymptotic value when $\epsilon\!=\!0$) vs. $\epsilon$. (b) Real part of the total electric field phasor obtained from 2D FDFD simulations, for an incoming $\hat{z}$-polarized Gaussian beam at normal incidence with respect to the air/nonlinear medium interface (represented by black dashed lines), with $\epsilon\!=\!0.1$ and $\epsilon\!=\!1$. The incoming electric field is normalized such that $\vert E(\omega)\vert_{(x,y)=(0,0)}\!=\!A_1(0)\!=\!5\!\times\!10^{8}$ [V/m]. In both panels, $\mu\!=\!1$ and $\chi^{(3)}\!=\!10^{-19}$ [(m/V)$^2$].}
\label{fig:AA}
\end{figure}

Let us for a moment step back and reflect on how these results on harmonic-generation processes can be connected to the optical Kerr effect and consider only the fundamental frequency $\omega$, in which case the nonlinear polarization can be written as
\begin{equation}
P_{NL}(\omega)=3\epsilon_0\chi^{(3)}(\omega;\omega,\omega,-\omega)\vert E(\omega)\vert^2E(\omega),\label{eq:A1}
\end{equation}
which yields, in the lossless case, an intensity-dependent refractive index
\begin{equation}
n=n_0+\Delta n=\sqrt{\mu}\sqrt{\epsilon+3\chi^{(3)}\vert E(\omega)\vert^2},\label{eq:A2}
\end{equation}
where $\Delta n$ is the nonlinear change in $n$ and is usually written as $\Delta n\!=\!\bar{n}_2\vert E(\omega)\vert^2$ or $\Delta n\!=\!n_2I$, with $\bar{n}_2\!=\!\frac{3\eta\chi^{(3)}}{4}$ and $n_2\!=\!\eta_0\eta\bar{n}_2\!=\!\frac{3\eta_0\eta^2\chi^{(3)}}{4}$ correct only to terms of order $I$ \cite{Boyd:2008:NOT:1817101,Reshef:17}: importantly, the same factor $\eta$ that shows up in the rate of increase of conversion efficiency in harmonic-generation processes now arises in the dependence of $\Delta n$ on $E$. The rate of phase change vs distance will therefore be $\frac{d\phi}{dz}\!=\!(n_0\!+\!\Delta n)\frac{\omega}{c}$, and thus the total nonlinear phase shift as measured in Z-scan experiments \cite{sheik1990sensitive} will be $\Delta n\frac{\omega}{c}L$, where $L$ is the length of the nonlinear medium; perhaps, though, the ratio of nonlinear to linear phase change, which we can approximate as $\frac{3\chi^{(3)}\vert E(\omega)\vert^2}{4\epsilon}\!=\!\frac{3\eta_0\sqrt{\mu}\chi^{(3)}I}{4\sqrt{\epsilon^3}}$, might represent a better (normalized) measure of the rate of nonlinear phase shift vs. distance. In any case, while it is true that the above-mentioned first-order correction leads to $\bar{n}_2\!\to\!\infty$ or $n_2\!\to\!\infty$ as $\epsilon\!\to\!0$ \cite{Reshef:17}, the fact remains that $\Delta n\!=\!\sqrt{\mu}\sqrt{\epsilon+3\chi^{(3)}\vert E(\omega)\vert^2}\!-\!\sqrt{\mu\epsilon}$, which is exact, tends to increase as $\epsilon$ decreases, up to the asymptotic value of $\sqrt{3\mu\chi^{(3)}}\vert E(\omega)\vert$ (this is seen in Fig.~\ref{fig:AA}a, where we compare the exact value of $\Delta n$ with its first-order approximation vs. $\epsilon$ when $\mu\!=\!1$). Consequently, the ratio of nonlinear to linear phase shift actually tends to $\infty$. Incidentally, note also that the relative error of the first-order approximation of $\Delta n$ is approximately constant with respect to $\mu$.

We implemented a nonlinear finite-difference frequency-domain (FDFD) \cite{hildebrand1968finite} full-wave solver so as to see the effect of this nonlinear phase shift numerically. Fig.~\ref{fig:AA}b shows the resulting electric field when, instead of a plane-wave, the excitation of our half-space problem is a normally-incident paraxial approximation of a Gaussian beam, with a beam waist radius of $4\lambda$, $\lambda$ being the vacuum wavelength. We choose the setup of this problem to be 2D with TM polarization, and compare the nonlinear results when $\epsilon\!=\!0.1$ and $\epsilon\!=\!1$, showing a larger beam distortion when $\epsilon\!=\!0.1$, as predicted by the theoretical on-axis increase of refractive index: $\Delta n_{\epsilon\!=\!0.1}\!=\!0.1021\!>\!\Delta n_{\epsilon\!=\!1}\!=\!0.0368$. 

\subsubsection{Two-Photon Absorption}

Analogous considerations apply if we consider the process of two-photon absorption \cite{Boyd:2008:NOT:1817101,agrawal2012nonlinear}, which we can describe also with Eq.~(\ref{eq:A1}) by making $\chi^{(3)}(\omega;\omega,\omega,-\omega)$ purely imaginary. If $\chi^{(3)}(\omega;\omega,\omega,-\omega)$ is generally complex, Eq.~(\ref{eq:A2}) becomes
\begin{equation}
n+i\frac{c\alpha}{2\omega}=\sqrt{\mu}\sqrt{\epsilon+3\chi^{(3)}\vert E(\omega)\vert^2},\label{eq:A3}
\end{equation}
where $\alpha\!=\!\alpha_0\!+\!\Delta\alpha\!=\!\alpha_0\!+\!\bar{\alpha}_2\vert E(\omega)\vert^2$ is the absorption coefficient, with $\bar{\alpha}_2$ the TPA coefficient.
Correct to first order in $I$ (and assuming no linear absorption for simplicity), we now have $\bar{\alpha}_2\!=\!\frac{3\eta\omega\text{Im}\left[\chi^{(3)}\right]}{2c}$, showing the same $\eta$-dependence seen in $\bar{n}_2$. Similarly as in the previous section, although $\Delta\alpha\!\to\!\infty$ (or $\bar{\alpha}_2\!\to\!\infty$ for that matter) as $\epsilon\!\to\!\infty$, the exact expression for $\Delta\alpha$ still does increase as $\epsilon$ decreases, until reaching the asymptote of value $\frac{2\sqrt{3\mu}\omega\text{Im}\left[\sqrt{\chi^{(3)}}\right]\vert E(\omega)\vert}{c}$.

\subsection{Imperfect Phase-Matching}
\begin{figure}[h]
\centering
\includegraphics[width=3.4in]{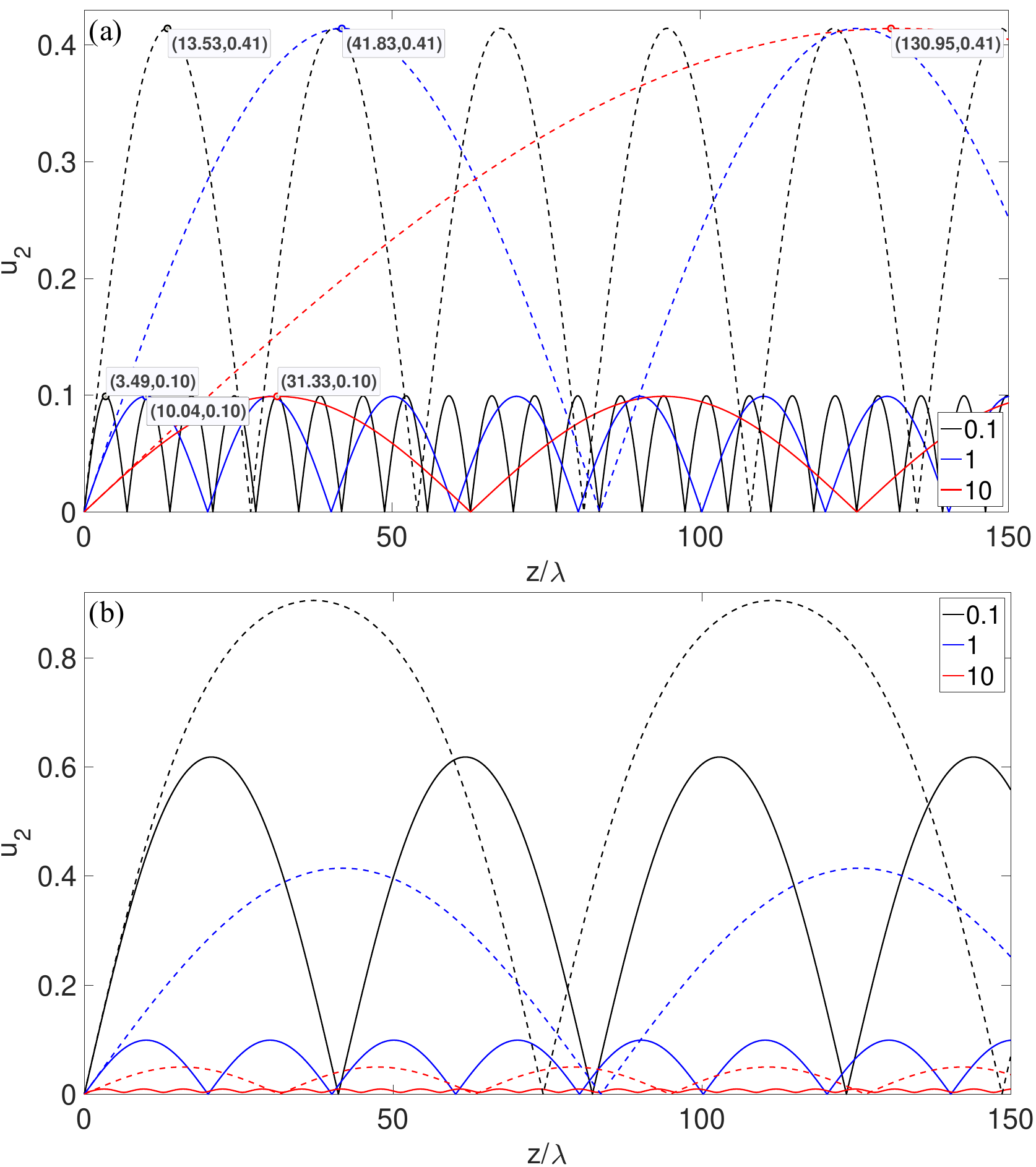}
\caption{Analytically-obtained (the expressions can be found in \cite{PhysRev.127.1918}) normalized field amplitudes of the second harmonic vs. normalized depth into the nonlinear medium, with $\epsilon_1$=0.1, 1 and 10, for $\epsilon_2$=$\epsilon_1+\Delta$ (panel (a), where the data cursors with XY-pairs $(z/\lambda,u_2)$ mark the first maximum of power conversion for each case) and $\epsilon_2$=$\epsilon_1(1+\Delta)$ (panel (b)). Solid lines: $\Delta=0.05$ ($\Delta s\!\approx\!-20$). Dashed lines: $\Delta=0.01$ ($\Delta s\!\approx\!-4$). The amplitude of the transmitted electric field at the entrance of the nonlinear medium is kept constant: $A_1(0)$=$5\!\times\!10^{8}$ [V/m].}
\label{fig:5}
\end{figure}

If the wavevectors are mismatched such that $\Delta s\! \neq\! 0$, the integration constant is now $\Gamma\! +\! \frac{\Delta s}{2}u_2^2(0)$ \cite{PhysRev.127.1918} and Eq.~(\ref{eq:7}) is generalized to
\begin{equation}
\zeta=\pm\frac{1}{2} \int_{u_2(0)}^{u_2}\frac{d(u_2^2)}{\sqrt{u_2^2(1-u_2^2)^2-\big[\Gamma-\frac{\Delta s}{2}\big(u_2^2-u_2^2(0)\big)\big]^2}}\label{eq:14}
\end{equation}
and $u_2$ now oscillates according to the solution in \cite{PhysRev.127.1918}, expressed in terms of the Jacobi elliptic function $sn()$ \cite{whittaker1996course}. For the initial condition of interest $u_2(0)$=$0$, $u_2$ will oscillate between 0 and $1/\left(\frac{\vert\Delta s\vert}{4}+\sqrt{1+(\frac{\vert\Delta s\vert}{4})^2}\right)$. The maximum of $u_2$ will hence increase with decreasing $\vert\Delta s\vert$, which for second-order polarization can be written as
\begin{equation}
\Delta s=\frac{2(n_1-n_2)}{\chi^{(2)}}\sqrt{\frac{2}{\eta_1^2\eta_2\eta_0I}}=\frac{2(n_1-n_2)}{\chi^{(2)}}\frac{1}{\sqrt{\eta_1\eta_2}\vert A_1(0)\vert}.\label{eq:15}
\end{equation}
Incidentally, note that $\Delta s$ as defined here will, in general, be a negative number due to Foster's reactance theoreom \cite{doi:10.1002/j.1538-7305.1924.tb01358.x}. If we assume there is no magnetic response, $\vert\Delta s\vert$ will be proportional to $(\sqrt{\epsilon_1}\! -\! \sqrt{\epsilon_2}){\sqrt[4]{\epsilon_1\epsilon_2}}/\vert A_1(0)\vert$. To study the dependence on $\epsilon_1$ and $\epsilon_2$ more fully, we assume that $\epsilon_2$=$\epsilon_1\!+\!\Delta$, we fix $\Delta$ and vary $\epsilon_1$: we find that 
$(\sqrt{\epsilon_1}\! -\! \sqrt{\epsilon_2}){\sqrt[4]{\epsilon_1\epsilon_2}}$ is practically constant (this is seen in Fig.~\ref{fig:5}a, where the maximum of power conversion is independent of $\epsilon_1$), yet $\vert\Delta k\vert\!\propto\!\vert\sqrt{\epsilon_1}\!-\! \sqrt{\epsilon_2}\vert$ decreases with $\epsilon_1$ (i.e., a smaller $\epsilon_1$ will render a larger $\vert\Delta k\vert$ but have practically no effect on $\vert\Delta s\vert$). With respect to intensity, nevertheless, $\vert\Delta s\vert\!\propto\!\sqrt[4]{\epsilon_1}$. It is paramount to realize, however, that $A_1(0)$ (or $I\!=\!I_1(0)$ for that matter) is referred to the inner side of the vacuum/nonlinear medium interface, so the transmission coefficient, which for normal incidence diminishes with $\epsilon_1$ as $\frac{2}{1+\sqrt{\epsilon_1}}$, plays an important role: conversion efficiency in the nonlinear medium still increases with decreasing $\epsilon$ when defined with respect to the incident intensity in vacuum. If we consider, alternatively, $\epsilon_2$=$\epsilon_1(1\! +\! \Delta)$, then $\vert\Delta s\vert\!\propto\!\epsilon_1$, so the maximum of power conversion grows with diminishing $\epsilon_1$, as depicted in Fig.~\ref{fig:5}b.

As expected, the behavior is the opposite if we consider the case in which $\epsilon_1\!=\!\epsilon_2\!=\!1$ and vary $\mu_{1,2}$. For $\mu_2$=$\mu_1\! +\! \Delta$, $\vert\Delta s\vert$ grows with diminishing $\mu_1$, whereas for $\mu_2$=$\mu_1(1\!+\!\Delta)$, $\frac{\sqrt{\mu_1}\! -\! \sqrt{\mu_2}}{\sqrt[4]{\mu_1\mu_2}}$ is a constant equal to $\frac{1-\sqrt{1+\Delta}}{\sqrt[4]{1\! +\! \Delta}}$ (with respect to $I$, $\vert\Delta s\vert\!\propto\!1/\sqrt[4]{\mu_1}$). The transmission coefficient for normal incidence will now grow with $\mu_1$ as $\frac{2\sqrt{\mu_1}}{1+\sqrt{\mu_1}}$.

In addition, although the oscillation period of $u_2(\zeta)$ decreases with increasing $\vert\Delta s\vert$ \cite{PhysRev.127.1918}, $\zeta$ is just a stretched/compressed version of z, so it is easy to see that $u_2(z)$, for fixed $\Delta s$, will see its period reduced as $\epsilon_1$ ($\mu_1$) decreases (increases). In other words, when $\Delta s$ is roughly constant with respect to $\epsilon_1$ ($\epsilon_2$=$\epsilon_1\!+\!\Delta$ and $\mu_1\!=\!\mu_2\!=\!1$) or $\mu_1$ ($\mu_2$=$\mu_1(1\!+\!\Delta)$ and $\epsilon_1\!=\!\epsilon_2\!=\!1$), and assuming that $A_1(0)$ is fixed, the first maximum of power conversion is found at a distance into the medium that roughly scales with $\sqrt{\epsilon_1/\mu_1}$, given that $z\!\propto\!\frac{\zeta}{\sqrt{\eta_1\eta_2}\vert A_1(0)\vert}\!\approx\!\frac{\zeta}{\eta_1\vert A_1(0)\vert}$ for sufficiently small $\Delta$. This is illustrated in Fig.~\ref{fig:5}a, where the data cursors mark the position of these maxima.

Having reached this point, it is imperative to note that our FDTD half-space problem in Figs.~\ref{fig:2}-\ref{fig:4} is not exactly described, at the inner face of the boundary, by the initial conditions assumed throughout the analytical derivations for propagation in a nonlinear medium. In actuality, weak higher harmonic waves are generated in reflection at the interface. Restricting the problem to second-harmonic generation, and neglecting the $\chi^{(2)}(\omega;2\omega,-\omega)$ process, the boundary value problem can be easily solved as in \cite{PhysRev.128.606}, whose generalization to account for (linear) magnetic permeability yields the following expression, restricted here to the simplified scenario of normal incidence, for the amplitude of the reflected electric field at the second-harmonic frequency:
\begin{equation}
A_2(0)=\frac{\sqrt{\mu_2}(\sqrt{\mu_1\epsilon_1}-\sqrt{\mu_2\epsilon_2})}{(\sqrt{\mu_2}+\sqrt{\epsilon_2})(-\mu_1\epsilon_1+\mu_2\epsilon_2)}\chi^{(2)}A_1^2(0).\label{eq:16}
\end{equation}
As for the second-harmonic wave transmitted into the nonlinear medium, it can be expressed as the superposition of a plane-wave with wavenumber $k_2$---general solution to the homogeneous wave equation---and a particular solution to the nonhomogeneous equation, in this case a plane-wave with wavenumber $2k_1$; or in more compact form:
\begin{equation}
E_2(z)=A_2(0)\left[1-\frac{\sqrt{\mu_2}(\sqrt{\mu_2}+\sqrt{\epsilon_2})}{\sqrt{\mu_1\epsilon_1}-\sqrt{\mu_2\epsilon_2}}(e^{i\Delta kz}-1)\right]e^{ik_2z}.\label{eq:17}
\end{equation}
It is clear that $E_2(0)$, obviously equal to the reflected wave's amplitude $A_2(0)$, can be used as initial condition for Eqs.~(\ref{eq:4a}),(\ref{eq:4b}), which do take into account the coupling of $E_2(z)$ into $E_1(z)$ described by $\chi^{(2)}(\omega;2\omega,-\omega)$. If $\mu_1$=$\mu_2$=1, and we calculate the limit when $\epsilon_1\!\to\!\epsilon_2$, the above expression has a simplified factor of $-\frac{1}{2(\epsilon_1+\sqrt{\epsilon_1})}$. Similarly, if $\epsilon_1$=$\epsilon_2$=1, the limit when $\mu_1\!\to\!\mu_2$ is $-\frac{1}{2(1+\sqrt{\mu_1})}$. That is, reducing $\epsilon$ not only increases conversion efficiency but also the amplitude of the reflected second-harmonic wave. Interestingly, though, increasing $\mu$ increases conversion efficiency but decreases nonlinear reflection. Going back to imperfectly matched phase velocities, it was stated before that for a fixed $\Delta$=$\epsilon_2$-$\epsilon_1$, the maximum of conversion efficiency is independent of $\epsilon_1$. On the contrary, the ratio $\frac{\sqrt{\epsilon_1}-\sqrt{\epsilon_1+\Delta}}{\Delta(1+\sqrt{\epsilon_1+\Delta})}$, and thus $u_2(0)$=$A_2(0)$, now decreases with $\epsilon_1$. If one realizes that $\theta(0)$=$\pi$ for real $\chi^{(2)}$, $u_2(0)$ is exactly the lower root of the denominator in Eq.~(\ref{eq:14}): without loss, $A_2(0)$ is a negative real number and, for very small $z$, the term $e^{i\Delta kz}\!-\!1$ is purely imaginary and grows linearly with $z$; if we match this initial condition at the interface with propagation in the bulk, we have $\phi_2(0)$=$\pi$ which, assuming $\phi_1(0)$=$0$, implies $u_2(0)$ is a minimum. Therefore reducing $\epsilon_1$ can raise the bounds of oscillation of $u_2(z)$.

\begin{figure}[h]
\centering
\includegraphics[width=3.4in]{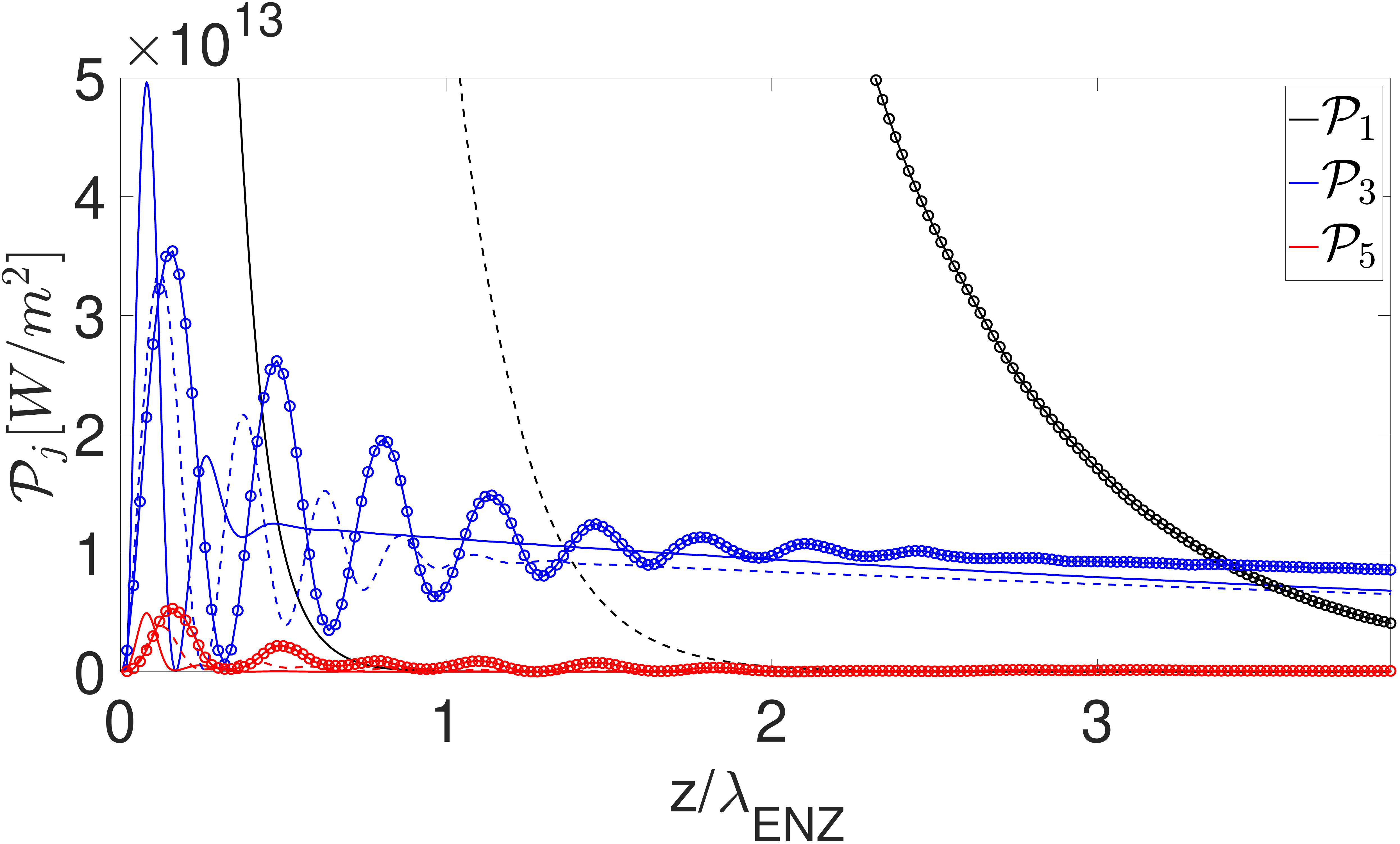}
\caption{FDTD-calculated time-averaged Poynting vector $\mathcal{P}_j $ for the first three harmonics vs. distance (normalized with respect to the ENZ wavelength), for different wavelengths of the fundamental, such that $\epsilon_1$=$-1.00+0.50i$, $0.00+0.35i$ and $1.00+0.22i$ (solid, dashed and marked lines, respectively). While the $\omega_1$ wave feels a metallic medium, the $\omega_3$ (and higher harmonics) wave undergoes much lower losses, which allows for propagation.}
\label{fig:6}
\end{figure}

The presence of losses in the nonlinear material substantially degrades power conversion. Yet it might be of interest to exploit the dispersion of the linear permittivity to our advantage by centering the fundamental harmonic at a frequency for which the material possesses metallic character but behaves essentially as a dielectric for higher harmonics. This transition region can be found in Drude-type plasmonic materials around the ENZ frequency. In Fig.~\ref{fig:6} the time-averaged Poynting vector is depicted vs. distance into the unbounded nonlinear half-space, in this case ITO, for $E_0$=$10^9$ [V/m] (incident intensity of $1.33\times10^{15}$ [W/m$^2$]).  The linear dielectric function of ITO is assumed to follow a Drude model with the parameters of \cite{Alam795}: free-electron plasma frequency $\omega_p$=$2.9719\times10^{15}$ [rad/s], collision frequency $\gamma$=$0.0468\omega_p$, and high-frequency permittivity $\epsilon_{\infty}$=3.8055. These constants fix the ENZ wavelength at 1240 nm. Three wavelengths are considered for the fundamental excitation: 1397, 1240 and 1065 nm, such that $\text{Re}\left\{\epsilon_1\right\}$=-1, 0 and 1, respectively (Table~\ref{table:1} lists all the complex permittivity values for the first three harmonics). A larger nonlinear susceptibility $\chi^{(3)}$=$2\!\times\!10^{-18}$ [m$^2$/V$^2$] is considered in this case to counteract losses. Fig.~\ref{fig:6} clearly shows how the third harmonic carries more power than the fundamental after a certain distance, as it experiences a much lower decay. 
\begin{table}[h]
\begin{center}
\begin{tabular}{ |c||c|c|c| } 
\hline
 & $\epsilon_1$ & $\epsilon_3$ & $\epsilon_5$ \\
\hline
Solid & $-1.00+0.50i$ & $+3.27+0.02i$ & $+3.61+0.00i$\\
Dashed & $+0.00+0.35i$ & $+3.38+0.01i$ & $+3.65+0.00i$\\
Marked & $+1.00+0.22i$ & $+3.49+0.00i$ & $+3.69+0.00i$\\
\hline
\end{tabular}
\end{center}
\caption{Complex values of the medium's linear permittivity at $\omega$, $3\omega$ and $5\omega$ for the three scenarios of Fig.~(\ref{fig:6}), with fundamental wavelengths 1397, 1240 and 1065 nm (solid, dashed and marked lines, respectively).}
\label{table:1}
\end{table}

\begin{figure}[h]
\centering
\includegraphics[width=3.4in]{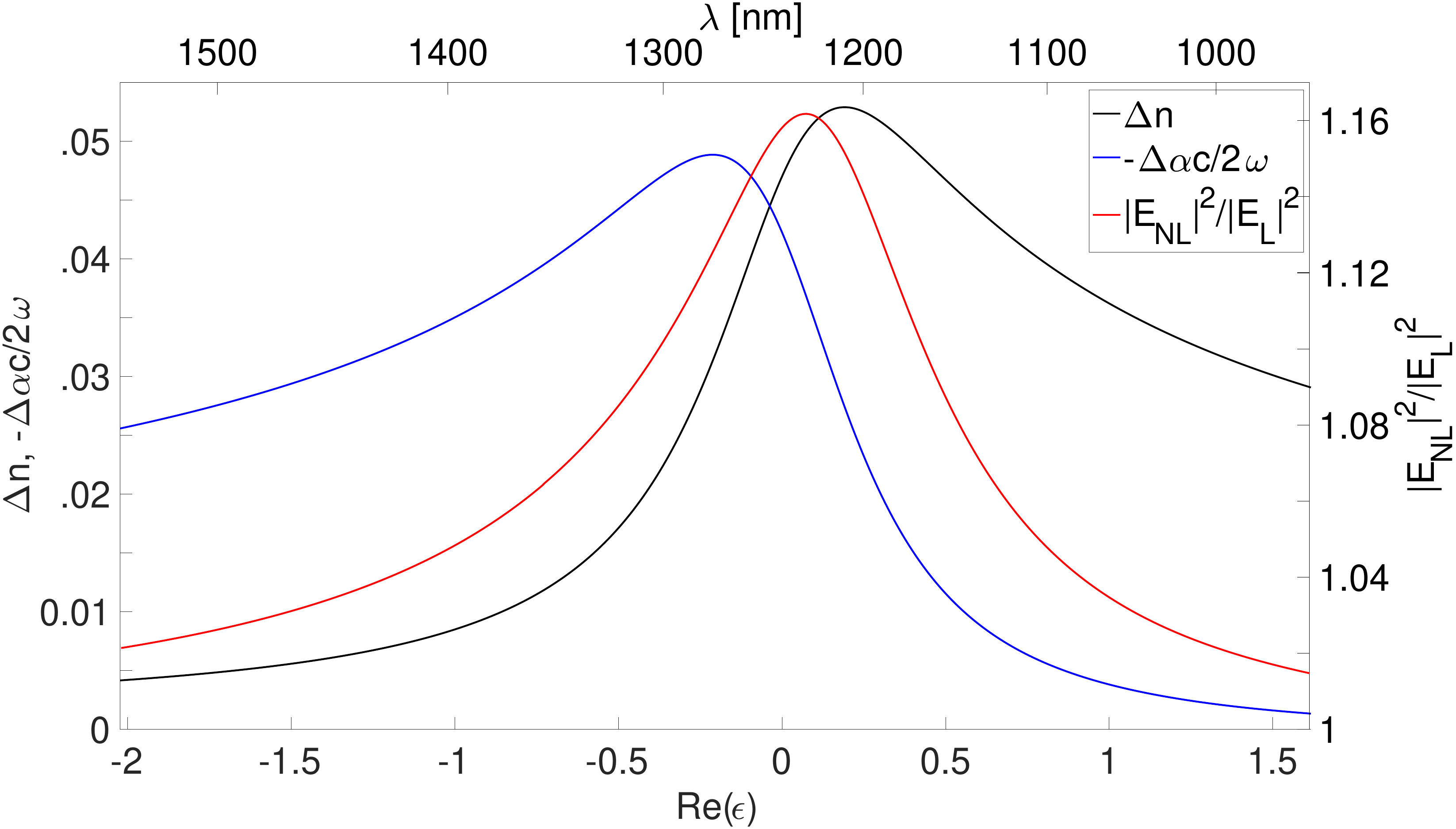}
\caption{1D FDFD simulations with the same setup as in Fig.~\ref{fig:6}, considering $\chi^{(3)}\!=\!10^{-19}$ [(m/V)$^2$] and an incident electric field of $\vert E(\omega)\vert\!=\!5\!\times\!10^{8}$ [V/m]. $\Delta n$ and $\Delta\alpha$ are depicted vs. $\text{Re}[\epsilon]$ following the Drude model of parameters indicated above, in the wavelength range of [940,1540] nm. We also perform linear simulations ($\chi^{(3)}\!=\!0$) and calculate the ratio of nonlinear/linear intensity at $z\!=\!4\lambda_{\text{ENZ}}$ (red curve in the plot, following a different ordinate axis as indicated on the right-side).}
\label{fig:8}
\end{figure}

Leaving harmomic generation aside, let us now restrict the discussion to $\chi^{(3)}(\omega;\omega,\omega,-\omega)$ processes only. If we calculate $\Delta n$ and $\Delta\alpha$ with the dielectric function of the Drude model and use the same $\chi^{(3)}$ as in Fig.~\ref{fig:AA}, we get, respectively, the black and blue curves in Fig.~\ref{fig:8}. One can see how the maxima of $\Delta n$ and $\Delta\alpha$ are slightly blue-shifted and red-shifted with respect to the ENZ frequency, respectively. Note also that the plot is showing $-\Delta \alpha$ (more precisely, normalized as the imaginary part of the complex refractive index), i.e., the Kerr effect is effectively reducing absorption loss. If we now perform monochromatic nonlinear FDFD simulations with the setup of Fig.~\ref{fig:6}, and measure the ratio of nonlinear-to-linear intensity $\frac{\vert E_{NL}(\omega)\vert^2}{\vert E_L(\omega)\vert^2}$ at a depth of four ENZ wavelengths, we obtain the red curve in Fig.~\ref{fig:8}, with a maximum of nonlinear response at $\lambda\!=\!1229$ nm, very close to the ENZ wavelength of 1240 nm. Importantly, these results are perfectly consistent with the experimental observations of an enhanced nonlinear response from ITO thin layers reported in \cite{Alam795,Alam2018}, where the origin of the nonlinearity is explained semi-classically with electron band theory: the laser induces a temperature rise of free electrons, which lowers the temperature-dependent electron chemical potential of ITO's nonparabolic conduction band and, in turn, reduces the plasma frequency $\omega_p$, thereby effectively increasing the dielectric function.

\section{Conclusions}

We have theoretically shown that the strength of the nonlinear response of a material tends to increase with an increasing linear permeability and/or a decreasing linear permittivity, according to a conversion length that decreases with an increasing relative impedance, under phase-matched conditions. We have also seen how, in considering the Kerr effect and TPA, this $\sqrt{\mu/\epsilon}$ dependence emerges in the nonlinear change of both the refractive index and the absorption coefficient. Moreover, if wave propagation is phase-mismatched, we have considered two scenarios (we herein restrict the notation to second-harmonic generation): if $\epsilon_2-\epsilon_1$ ($\mu_2-\mu_1$) is kept constant, the maximum of power conversion does not vary with $\epsilon_1$ (increases with $\mu_1$) and the oscillation period increases with $\epsilon_1$ (increases with $\mu_1$); if $\epsilon_2/\epsilon_1$ ($\mu_2/\mu_1$) is kept constant, the maximum of power conversion decreases with $\epsilon_1$ (does not vary with $\mu_1$) and the oscillation period decreases with $\epsilon_1$ (decreases with $\mu_1$). Consequently, either the oscillation amplitude of power conversion tends to increase with increasing $\mu$ and/or decreasing $\epsilon$, or else this amplitude stays constant with respect to $\mu$ and $\epsilon$, but with an oscillation period that decreases with increasing $\mu$ and/or decreasing $\epsilon$.

The behavior described here is consistent with previous experimental measurements of unusually large nonlinear phase shifts of ENZ materials \cite{Alam795,PhysRevLett.116.233901}, and yet proves that a stronger nonlinear response---restricted in this paper to low-order harmonic-generation processes, Kerr effect and TPA---does not necessarily require a larger nonlinear susceptibility, but can rather be traced back to the relative strengths of the electric and magnetic fields, quantified through the relative impedance. This work thus shows that, even at a fixed value of the nonlinear susceptibility, one can obtain a larger overall nonlinear response, such as an increased conversion efficiency for low-order harmonic generation, by choosing situations such that one or more of the interacting frequencies lies in an ENZ region of the nonlinear material.

\begin{acknowledgments}
This work is supported in part by the Defense Advanced Research Projects Agency (DARPA) Defense Sciences Office (DSO) Nascent Light-Matter Interaction program under grant number W911NF-18-0369.
\end{acknowledgments}

\nocite{*}

\bibliography{apssamp}

\end{document}